\documentclass[final,3p,times,authoryear]{elsarticle}

\usepackage{amssymb}
\usepackage{amsmath}
\usepackage{amsthm}
\usepackage{hyperref}
\usepackage{float}
\usepackage{cleveref}
\usepackage{tikz}
\usepackage{booktabs}
\usepackage{multirow}
\usetikzlibrary{decorations.pathreplacing,arrows.meta,calc}

\theoremstyle{definition}
\newtheorem{defn}{Definition}[]
\theoremstyle{plain}
\newtheorem{prop}{Proposition}[]

\journal{Finance Research Letter}

\begin{document}
\global\long\def\bC{\mathbb{C}}%

\global\long\def\bE{\mathbb{E}}%

\global\long\def\bF{\mathbb{F}}%

\global\long\def\bK{\mathbb{K}}%

\global\long\def\bN{\mathbb{N}}%

\global\long\def\bP{\mathbb{P}}%

\global\long\def\bQ{\mathbb{Q}}%

\global\long\def\bR{\mathbb{R}}%

\global\long\def\bT{\mathbb{T}}%

\global\long\def\bZ{\mathbb{Z}}%

\global\long\def\cA{\mathcal{A}}%

\global\long\def\cB{\mathcal{B}}%

\global\long\def\cC{\mathcal{C}}%

\global\long\def\cD{\mathcal{D}}%

\global\long\def\cE{\mathcal{E}}%

\global\long\def\cF{\mathcal{F}}%

\global\long\def\cG{\mathcal{G}}%

\global\long\def\cH{\mathcal{H}}%

\global\long\def\cI{\mathcal{I}}%

\global\long\def\cJ{\mathcal{J}}%

\global\long\def\cK{\mathcal{K}}%

\global\long\def\cL{\mathcal{L}}%

\global\long\def\cLp#1{\mathcal{L}^{#1}}%

\global\long\def\cLpp#1{\mathcal{L}_{+}^{#1}}%

\global\long\def\cLsimp{\mathcal{L}_{simp}^{0}}%

\global\long\def\cM{\mathcal{M}}%

\global\long\def\cN{\mathcal{N}}%

\global\long\def\cO{\mathcal{O}}%

\global\long\def\cP{\mathcal{P}}%

\global\long\def\cR{\mathcal{R}}%

\global\long\def\cS{\mathcal{S}}%

\global\long\def\cT{\mathcal{T}}%

\global\long\def\cU{\mathcal{U}}%

\global\long\def\cV{\mathcal{V}}%

\global\long\def\cW{\mathcal{W}}%

\global\long\def\cY{\mathcal{Y}}%

\global\long\def\cZ{\mathcal{Z}}%

\global\long\def\fq{\mathscr{\mathfrak{q}}}%

\global\long\def\fH{\mathscr{\mathfrak{H}}}%

\global\long\def\sD{\mathscr{D}}%

\global\long\def\sH{\mathscr{H}}%

\global\long\def\sK{\mathscr{K}}%

\global\long\def\sF{\mathscr{F}}%

\global\long\def\norm#1{\left\Vert #1\right\Vert }%

\global\long\def\np#1#2{\left\Vert #1\right\Vert _{#2}}%

\global\long\def\nlp#1#2{\left\Vert #1\right\Vert _{L^{#2}}}%

\global\long\def\abs#1{\left|#1\right|}%

\global\long\def\inv#1{#1^{-1}}%

\global\long\def\adjoint#1{#1^{*}}%

\global\long\def\annihilator#1{#1^{\circ}}%

\global\long\def\annihilatee#1{#1^{\perp}}%

\global\long\def\unaryop#1{#1\left(\cdot\right)}%

\global\long\def\binaryop#1{#1\left(\cdot,\cdot\right)}%

\global\long\def\comp#1#2{#1\circ#2}%

\global\long\def\converge#1{\overset{#1}{\joinrel\longrightarrow}}%

\global\long\def\define{\triangleq}%

\global\long\def\enum#1#2{\left\{  #1_{1},\dots,#1_{#2}\right\}  }%

\global\long\def\enumvec#1#2{\left(#1_{1},\dots,#1_{#2}\right)}%

\global\long\def\enuminf#1{\left\{  #1_{1},#1_{2}\dots\right\}  }%

\global\long\def\equivalent{\Longleftrightarrow}%

\global\long\def\substitute#1{\overset{#1}{\joinrel===}}%

\global\long\def\tensor{\otimes}%

\global\long\def\liminf#1{\underset{#1}{\operatorname{lim\,inf}}}%

\global\long\def\limsup#1{\underset{#1}{\operatorname{lim\,sup}}}%

\global\long\def\essinf#1{\underset{#1}{\operatorname{ess\,inf}}}%

\global\long\def\esssup#1{\underset{#1}{\operatorname{ess\,sup}}}%

\global\long\def\sgn{\operatorname{sgn}}%

\global\long\def\spanset{\operatorname{span}}%

\global\long\def\Null{\operatorname{Null}}%

\global\long\def\Range{\operatorname{Range}}%

\global\long\def\io{\operatorname{i.o.}}%

\global\long\def\ae{\operatorname{a.e.}}%

\global\long\def\as{\operatorname{a.s.}}%

\global\long\def\d#1{\operatorname{d}#1}%

\global\long\def\D#1{\operatorname{D}#1}%

\global\long\def\Db#1{\operatorname{D}\left[#1\right]}%

\global\long\def\cov{\operatorname{cov}}%

\global\long\def\supp{\operatorname{supp}}%

\begin{frontmatter}

\title{Reinforcement Learning for Option Hedging: Static Implied-Volatility Fit versus Shortfall-Aware Performance}

\author[label1]{Ziheng Chen}
\ead{stokes615@utexas.edu}
\affiliation[label1]{
organization={Department of Mathematics, University of Texas at Austin},
addressline={2515 Speedway}, 
city={Austin},
state={TX},
country={USA}
}

\author[label2]{Minxuan Hu}
\ead{mh2229@cornell.edu}
\affiliation[label2]{
organization={Cornell Ann S. Bowers College of Computing and Information Science, Cornell University},
addressline={105 Campus Rd},
city={Ithaca},
state={NY},
country={USA}
}

\author[label3]{Jiayu Yi}
\ead{sophiayi97@gmail.com}
\affiliation[label3]{
organization={School of Social Sciences, Nanyang Technological University},
addressline={48 Nanyang Avenue},
country={Singapore}
}

\author[label4]{Wenxi Sun}
\ead{wsun41@alumni.jh.edu}
\affiliation[label4]{
organization={Krieger School of Arts and Sciences, Johns Hopkins University},
addressline={3400 N Charles St},
city={Baltimore},
state={MD},
country={USA}
}

\begin{abstract}
We extend the Q-learner in Black-Scholes (QLBS) framework by incorporating risk aversion and trading costs, and propose a novel Replication Learning of Option Pricing  (RLOP) approach.
Both methods are fully compatible with standard reinforcement learning algorithms and operate under market frictions.
Using SPY and XOP option data, we evaluate performance along static and dynamic dimensions.
Adaptive-QLBS achieves higher static pricing accuracy in implied volatility space, while RLOP delivers superior dynamic hedging performance by reducing shortfall probability.
These results highlight the importance of evaluating option pricing models beyond static fit, emphasizing realized hedging outcomes.
\end{abstract}



\end{frontmatter}

\makeatother

\section{Introduction}

Option pricing and hedging remain core challenges in quantitative finance. 
 \cite{black1973pricing} provide a foundational option valuation model where perfect replication is achieved under frictionless markets with continuous-time trading, and subsequent research has extended BSM to adapt to more complex market realities (\cite{fan2022empirical,li2025analytic,golbabai2013superconvergence}). Nevertheless, real-world markets involve transaction costs and discrete-time trading.

In this setting, reinforcement learning (RL) provides a more flexible framework for optimizing hedging strategies under transaction costs and tail-risk management, while also supporting robustness and stress testing via synthetic market scenarios (\cite{stutz2025jdapp}). \cite{halperin2020qlbs,halperin2019qlbs}
introduced the Q-Learner in Black-Scholes (QLBS) framework to unify option pricing and hedging into a discrete-time Markov Decision Process (MDP), while ignoring transaction costs. 
Subsequently,  \cite{buehler2019deep} introduce the ``Deep Hedging'' framework, using neural networks to optimize hedging strategies under convex risk measures and market frictions. However, these methodologies primarily focus on symmetric or expectation-based risk metrics that capture large losses. According to \cite{follmer2000efficient} , when perfect replication is impracticable due to market frictions, hedging should minimize ``shortfall risk''-the probability of underperforming the option payoff.

To bridge methodological gaps, this paper addresses the decoupling between pricing-model calibration and real-world hedging performance. We shift the hedging objective from traditional error-minimization to shortfall-probability optimization to better manage transaction costs and tail risks. Using a neural-network-based RL agent within modified QLBS and novel RLOP frameworks, we compare optimal strategies against the Black-Scholes benchmark, which explicitly incorporates risk aversion and market frictions. The study investigates that under high transaction costs, the two RL-based approaches exhibit fundamentally divergent hedging behaviors. the modified QLBS agent optimizes for cost-aware stability, while the RLOP agent mitigates margin pressure and liquidity demand by systematically reducing exposure during extreme stress.

This paper delivers three primary research contributions: (1) This paper extends the QLBS framework by operationalizing its latter computational stage and embedding shortfall probability into its reward structure, resolving the decoupling between IV-fitting and hedging performance and transitioning the agent from simple loss-minimization to a survival-centric hedging strategy. (2) The novel RLOP model is introduced for superior tail-risk resilience and computational efficiency, which prioritizes minimizing hedging failure frequency over loss magnitude, as empirically validated during the COVID-19 crash. (3) This paper offers a selection framework through a bidirectional computational architecture, where the backward-operating QLBS acts as a cost-aware stabilizer for volatile assets, and the forward-calculating RLOP serves capital-constrained desks by minimizing turnover and tail losses, together yielding more robust hedging outputs.

The remainder of this paper is structured as follows: Section 2 and 3 introduce the modified QLBS model and the novel RLOP model, Section 4 and 5 present their empirical analysis under varying risk and cost scenarios, and Section 6 concludes this paper.

\section{Replication Pricing and Reinforcement Learning Framework}
Replication-based pricing constructs a self-financing portfolio whose terminal value matches the option payoff.
In the classical framework by \cite{black1973pricing}, this is achieved through dynamic rebalancing of a hedge portfolio.
Given a price process $\left\{ S_{t}\right\}$
adapted to a filtration $\left\{ \cF_{t}\right\} $, the portfolio consists of $u_t$ units of the underlying asset and a risk-free deposit $B_{t}$, with value
$
\Pi_{t}:=u_{t}S_{t}+B_{t}.
$
The self-financing condition requires that rebalancing does not inject or withdraw capital, yielding
\begin{equation}
u_{t}S_{t+1}+e^{r\Delta t}B_{t}=u_{t+1}S_{t+1}+B_{t+1} + \text{TC}(u_{t+1}-u_t, S_{t+1}) \label{eq:self-financing}
\end{equation}
where $r$ denotes the risk-free rate and $\text{TC}(\cdot)$ represents the transaction cost.
\Cref{eq:self-financing} governs the portfolio evolution and determines the capital required to sustain trading.
Throughout, the underlying price is assumed to follow a geometric Brownian motion
$ \d S_t=\mu \,S_t \d{t} + \sigma \,S_t \d{W_t} $ 
with drift $\mu$, volatility $\sigma$, and Wiener process $W_t$.

Reinforcement learning (RL) formulates sequential decision problems via a Markov decision process (MDP).
Per \cite{sutton2018reinforcement,bertsekas2019reinforcement},
an MDP is specified by a state space $\mathcal S$, an action space $\mathcal A$, a transition kernel $p(\cdot \mid s,a)$, and a reward function $R(s,a)$. 
A policy $\pi$ maps the current state to an action; in our setting, the policy corresponds to a hedging rule that updates the position over time.
Motivated by the Girsanov transform \cite{liptser2013statistics}, we cast option replication as an MDP with state $(t,X_t)$, where the normalized price process is
$
X_t := -\Bigl(\mu - \tfrac{\sigma^2}{2}\Bigr)t + \log S_t .
$
The action $a_t$ represents the hedge position selected from the normalized input $X_t$.
The transition kernel $p$ (and hence the realized rewards) depends on the model construction (e.g., QLBS versus RLOP as discussed in later paragraphs). 
To distinguish representations, we use $a_t$ for the hedge chosen as a function of $X_t$, while $u_t$ denotes the hedge position when expressed on the original price scale $S_t$.

As established by \cite{halperin2020qlbs}, the price from the Black-Scholes model can be characterized in discrete time as the expected value of the replicating portfolio $\Pi_t$.
This insight leads to the control problem where the fair option price is the maximum of the state value function
$
\tilde{V}_{t}^{\pi}\left(X_{t}\right)=\bE^{\pi}_t\left[-\Pi_{t}\left(X_{t}\right)-\lambda\sum_{\tau=t}^{T}e^{-r\left(\tau-t\right)}\text{Var}_t\left[\Pi_{\tau}\left(X_{\tau}\right)\right]\right]
$
where $\lambda$ is the risk-aversion coefficient and $\bE_t$ is conditional expectation under $\cF_t$.
Under this formulation, the option price equals the negative of the optimal value function.

\cite{halperin2020qlbs} derives a closed-form maximizer of 
$\tilde{V}_{t}^{\pi}$
by exploiting its quadratic mean-variance structure,
which is feasible when the correlation between  $\Pi_{t+1}$ and $\Delta S$ is known.
However, this approach does not generalize beyond the stylized setting.
For a given payoff function $h$, the portfolio value $\Pi_t$ that satisfies the terminal condition $\Pi_T = h(S_T)$ is typically non-adapted under the self-financing condition (\cref{eq:self-financing}), making direct application of standard RL algorithms nontrivial.

\section{Two Reinforcement Learning Paradigms for Option Pricing}

To overcome the limitations of the original QLBS formulation, two complementary strategies have been proposed. 
The first approach extends the existing QLBS framework by redefining the value function $V_t^\pi$ as an $\{\cF_t\}$-adapted process.
The second approach adopts a conceptually different perspective by constructing an adaptive portfolio value process and defining the reward in terms of the terminal hedging tracking error. 
Both approaches explicitly account for transaction costs and are compatible with both value-based and policy-based reinforcement learning algorithms.

\subsection{Adaptive QLBS: A Backward Value-Based RL Framework}



\begin{defn}
\label{def:qlbs}
Let $d_T(t):=\left(1-\frac{t}{T}\right)$ and $\gamma:=e^{-r \Delta t}$.
The proposed value function reads
\begin{align}
V_{t}^{\pi}\left(X_{t}\right)&:=\bE_{t}^{\pi}\left[-d_T(t)\Pi_{t}\left(X_{t}\right)-\lambda\sum_{\tau=t}^{T}\gamma^{\tau-t}\sqrt{\text{Var}\left[\Pi_{\tau}\left(X_{\tau}\right)\right]}\right],\label{eq:qlbs-modified-V}
\end{align}
with the reward function $R_{t+1}\left(X_{t},a_{t}\right):=V_t^\pi(X_t)-\bE_{t}^{\pi} V_{t+1}^\pi(X_{t+1})$ where $X_{t+1}$ is implicitly determined by $X_t,a_t$ and the self-financing condition \cref{eq:self-financing}
\end{defn}


Our modifications are twofold:
we introduce a diminishing factor $d_T(t)$ that weights the portfolio term from $1$ at $t=0$ to $0$ at $t=T$, smoothing the contribution of the terminal payoff,
and we replace variance terms by their square roots to obtain a dimensionless and numerically more stable value estimate.
When transaction costs are included, the portfolio value process $\Pi_{t}$ is computed backward using the self-financing condition \cref{eq:self-financing}.
A schematic illustration of the adaptive-QLBS model is provided in \cref{fig:qlbs-illustration}.


\begin{figure}[htbp]
    \centering
    \begin{minipage}{0.39\textwidth}
        \centering
        \begin{tikzpicture}[x=0.8cm,y=0.8cm, line cap=round, line join=round]

  \coordinate (O)   at (2,0); 
  \coordinate (t0)  at (4,0); \coordinate (t0a)  at (4,2.5);
  \coordinate (T)   at (7.5,0);
  \coordinate (tau) at (7.65,0);

  \draw[thick,dotted] (O) -- (t0);
  \draw[thick] (t0) -- (T);

  \node[below] at (O) {$0$};
  \node[below]      at (t0) {$t$};
  \node[below]      at (T) {$T$};

  \fill (t0) circle (2.2pt);


  \node[above] at ($(7.25,1.8)$) {$S_T$};
  \draw[color=gray] (t0a) --
  ++(0.055,0.117) -- ++(0.055,0.074) -- ++(0.055,0.149) -- ++(0.055,0.151) -- ++(0.055,0.152) -- ++(0.055,0.084) -- ++(0.055,0.018) -- ++(0.055,0.056) -- ++(0.055,0.105) -- ++(0.055,0.045) -- ++(0.055,0.083) -- ++(0.055,0.094) -- ++(0.055,0.085) -- ++(0.055,-0.047) -- ++(0.055,0.020) -- ++(0.055,0.103) -- ++(0.055,0.035) -- ++(0.055,0.006) -- ++(0.055,0.015) -- ++(0.055,0.013) -- ++(0.055,0.089) -- ++(0.055,-0.042) -- ++(0.055,0.073) -- ++(0.055,0.009) -- ++(0.055,0.006) -- ++(0.055,0.007) -- ++(0.055,0.054) -- ++(0.055,0.057) -- ++(0.055,0.050) -- ++(0.055,-0.035) -- ++(0.055,0.075) -- ++(0.055,0.051) -- ++(0.055,0.046) -- ++(0.055,-0.031) -- ++(0.055,0.088) -- ++(0.055,0.047) -- ++(0.055,0.019) -- ++(0.055,0.099) -- ++(0.055,-0.010) -- ++(0.055,0.073) -- ++(0.055,0.006) -- ++(0.055,-0.072) -- ++(0.055,0.027) -- ++(0.055,0.074) -- ++(0.055,-0.008) -- ++(0.055,-0.019) -- ++(0.055,-0.064) -- ++(0.055,0.020) -- ++(0.055,-0.035) -- ++(0.055,-0.007) -- ++(0.055,0.022) -- ++(0.055,-0.052) -- ++(0.055,0.058) -- ++(0.055,-0.052) -- ++(0.055,-0.032) -- ++(0.055,0.067) -- ++(0.055,-0.109) -- ++(0.055,0.025) -- ++(0.055,-0.021) -- ++(0.055,0.065) -- ++(0.055,0.026) -- ++(0.055,0.067) -- ++(0.055,-0.016) ;
  \draw[color=gray] (t0a) --
  ++(0.055,0.049) -- ++(0.055,0.082) -- ++(0.055,0.053) -- ++(0.055,0.020) -- ++(0.055,0.061) -- ++(0.055,0.102) -- ++(0.055,0.079) -- ++(0.055,-0.002) -- ++(0.055,-0.029) -- ++(0.055,0.003) -- ++(0.055,0.034) -- ++(0.055,-0.079) -- ++(0.055,0.024) -- ++(0.055,-0.025) -- ++(0.055,0.000) -- ++(0.055,0.009) -- ++(0.055,0.019) -- ++(0.055,0.052) -- ++(0.055,0.079) -- ++(0.055,0.020) -- ++(0.055,0.089) -- ++(0.055,-0.011) -- ++(0.055,0.037) -- ++(0.055,0.061) -- ++(0.055,0.019) -- ++(0.055,-0.021) -- ++(0.055,-0.028) -- ++(0.055,-0.005) -- ++(0.055,0.027) -- ++(0.055,-0.032) -- ++(0.055,0.007) -- ++(0.055,0.009) -- ++(0.055,0.042) -- ++(0.055,0.024) -- ++(0.055,0.029) -- ++(0.055,-0.020) -- ++(0.055,0.006) -- ++(0.055,0.049) -- ++(0.055,0.079) -- ++(0.055,-0.054) -- ++(0.055,0.079) -- ++(0.055,0.067) -- ++(0.055,0.036) -- ++(0.055,0.010) -- ++(0.055,-0.018) -- ++(0.055,0.067) -- ++(0.055,0.086) -- ++(0.055,0.074) -- ++(0.055,0.047) -- ++(0.055,-0.000) -- ++(0.055,-0.074) -- ++(0.055,-0.013) -- ++(0.055,0.019) -- ++(0.055,-0.074) -- ++(0.055,0.009) -- ++(0.055,0.011) -- ++(0.055,0.022) -- ++(0.055,-0.067) -- ++(0.055,-0.039) -- ++(0.055,-0.026) -- ++(0.055,0.026) -- ++(0.055,0.067) -- ++(0.055,-0.016) ;
  \draw[thick] (t0a) --
  ++(0.055,-0.009) -- ++(0.055,0.007) -- ++(0.055,-0.004) -- ++(0.055,-0.056) -- ++(0.055,0.029) -- ++(0.055,-0.023) -- ++(0.055,0.013) -- ++(0.055,-0.030) -- ++(0.055,0.054) -- ++(0.055,-0.011) -- ++(0.055,0.011) -- ++(0.055,-0.004) -- ++(0.055,0.023) -- ++(0.055,-0.002) -- ++(0.055,0.004) -- ++(0.055,0.038) -- ++(0.055,0.009) -- ++(0.055,0.021) -- ++(0.055,-0.024) -- ++(0.055,0.021) -- ++(0.055,-0.030) -- ++(0.055,0.020) -- ++(0.055,-0.045) -- ++(0.055,-0.050) -- ++(0.055,-0.009) -- ++(0.055,0.045) -- ++(0.055,0.008) -- ++(0.055,-0.015) -- ++(0.055,-0.004) -- ++(0.055,-0.065) -- ++(0.055,0.017) -- ++(0.055,-0.029) -- ++(0.055,0.022) -- ++(0.055,0.057) -- ++(0.055,0.005) -- ++(0.055,-0.007) -- ++(0.055,0.029) -- ++(0.055,0.010) -- ++(0.055,-0.012) -- ++(0.055,-0.031) -- ++(0.055,-0.043) -- ++(0.055,0.048) -- ++(0.055,0.019) -- ++(0.055,0.061) -- ++(0.055,-0.011) -- ++(0.055,-0.032) -- ++(0.055,0.030) -- ++(0.055,-0.026) -- ++(0.055,-0.058) -- ++(0.055,0.049) -- ++(0.055,-0.026) -- ++(0.055,0.014) -- ++(0.055,-0.019) -- ++(0.055,0.037) -- ++(0.055,-0.025) -- ++(0.055,0.009) -- ++(0.055,0.005) -- ++(0.055,-0.005) -- ++(0.055,0.044) -- ++(0.055,0.026) -- ++(0.055,0.026) -- ++(0.055,0.067);
  \draw[color=gray] (t0a) --
  ++(0.055,-0.127) -- ++(0.055,0.040) -- ++(0.055,-0.173) -- ++(0.055,-0.052) -- ++(0.055,-0.075) -- ++(0.055,-0.066) -- ++(0.055,-0.064) -- ++(0.055,-0.034) -- ++(0.055,0.024) -- ++(0.055,0.034) -- ++(0.055,-0.066) -- ++(0.055,-0.038) -- ++(0.055,-0.038) -- ++(0.055,-0.013) -- ++(0.055,-0.014) -- ++(0.055,0.023) -- ++(0.055,-0.013) -- ++(0.055,0.066) -- ++(0.055,-0.027) -- ++(0.055,-0.012) -- ++(0.055,-0.045) -- ++(0.055,-0.074) -- ++(0.055,-0.084) -- ++(0.055,0.018) -- ++(0.055,-0.079) -- ++(0.055,-0.042) -- ++(0.055,-0.057) -- ++(0.055,-0.009) -- ++(0.055,0.005) -- ++(0.055,-0.062) -- ++(0.055,-0.036) -- ++(0.055,0.014) -- ++(0.055,-0.038) -- ++(0.055,-0.035) -- ++(0.055,-0.006) -- ++(0.055,-0.051) -- ++(0.055,0.176) -- ++(0.055,0.042) -- ++(0.055,-0.060) -- ++(0.055,0.025) -- ++(0.055,0.002) -- ++(0.055,-0.044) -- ++(0.055,0.036) -- ++(0.055,0.027) -- ++(0.055,-0.097) -- ++(0.055,0.029) -- ++(0.055,-0.003) -- ++(0.055,-0.066) -- ++(0.055,-0.080) -- ++(0.055,-0.048) -- ++(0.055,-0.035) -- ++(0.055,0.035) -- ++(0.055,0.086) -- ++(0.055,0.015) -- ++(0.055,0.064) -- ++(0.055,0.049) -- ++(0.055,0.046) -- ++(0.055,-0.136) -- ++(0.055,0.072) -- ++(0.055,0.016)  -- ++(0.055,0.026) -- ++(0.055,0.067);
  \draw[color=gray] (t0a) --
  ++(0.055,-0.032) -- ++(0.055,-0.053) -- ++(0.055,-0.214) -- ++(0.055,-0.108) -- ++(0.055,-0.126) -- ++(0.055,-0.151) -- ++(0.055,-0.081) -- ++(0.055,-0.168) -- ++(0.055,-0.102) -- ++(0.055,-0.070) -- ++(0.055,-0.016) -- ++(0.055,-0.028) -- ++(0.055,-0.087) -- ++(0.055,-0.023) -- ++(0.055,-0.027) -- ++(0.055,-0.027) -- ++(0.055,-0.137) -- ++(0.055,0.060) -- ++(0.055,-0.097) -- ++(0.055,-0.146) -- ++(0.055,-0.040) -- ++(0.055,-0.066) -- ++(0.055,-0.011) -- ++(0.055,0.046) -- ++(0.055,-0.061) -- ++(0.055,-0.062) -- ++(0.055,0.002) -- ++(0.055,0.001) -- ++(0.055,0.013) -- ++(0.055,0.071) -- ++(0.055,0.059) -- ++(0.055,0.001) -- ++(0.055,-0.013) -- ++(0.055,0.016) -- ++(0.055,0.022) -- ++(0.055,0.031) -- ++(0.055,0.022) -- ++(0.055,-0.054) -- ++(0.055,0.021) -- ++(0.055,-0.017) -- ++(0.055,-0.012) -- ++(0.055,-0.009) -- ++(0.055,-0.011) -- ++(0.055,0.031) -- ++(0.055,0.056) -- ++(0.055,-0.114) -- ++(0.055,-0.057) -- ++(0.055,0.092) -- ++(0.055,-0.048) -- ++(0.055,0.058) -- ++(0.055,-0.019) -- ++(0.055,0.110) -- ++(0.055,0.034) -- ++(0.055,-0.016) -- ++(0.055,-0.015) -- ++(0.055,-0.041) -- ++(0.055,-0.065) -- ++(0.055,-0.102) -- ++(0.055,-0.056) -- ++(0.055,-0.058) -- ++(0.055,-0.028) -- ++(0.055,-0.034) ;
  \draw[thick,dotted] (t0a) --
  ++(-0.050,0.049) -- ++(-0.050,-0.165) -- ++(-0.050,0.030) -- ++(-0.050,0.037) -- ++(-0.050,0.257) -- ++(-0.050,-0.076) -- ++(-0.050,-0.015) -- ++(-0.050,0.020) -- ++(-0.050,-0.223) -- ++(-0.050,-0.071) -- ++(-0.050,0.079) -- ++(-0.050,-0.094) -- ++(-0.050,-0.127) -- ++(-0.050,0.003) -- ++(-0.050,0.103) -- ++(-0.050,-0.191) -- ++(-0.050,-0.141) -- ++(-0.050,-0.083) -- ++(-0.050,-0.069) -- ++(-0.050,0.002) -- ++(-0.050,-0.159) -- ++(-0.050,0.109) -- ++(-0.050,0.158) -- ++(-0.050,0.006) -- ++(-0.050,-0.092) -- ++(-0.050,0.014);

  \draw[-{Stealth[length=2mm,width=1.5mm]}] ($(t0)+(0,1.5)$) -- ($(t0)+(0,2)$);

  \draw[dashed,thick] ($(tau)+(0,0.3)$) -- ($(tau)+(0,4.8)$);


  \node[align=center] at ($(t0a)+(-0.8,1.4)$)
    {$\text{Var}\,\Pi_t$ \\under\\ \cref{eq:self-financing}};

  \node[align=center,above] at ($(t0a)+(-0,-2.35)$) {hedge $u_t$ \\ state $(t,S_t)$};

  \node[align=right] at ($(tau)+(-1,5.5)$)
    {terminal \\condition $h$};

\end{tikzpicture}
        \caption{The adaptive-QLBS method takes a backward, value-based approach.}
        \label{fig:qlbs-illustration}
    \end{minipage}
    \hfill
    \begin{minipage}{0.59\textwidth}
        \centering
        \begin{tikzpicture}[x=0.8cm,y=0.8cm,>=Stealth]
  \tikzset{
    tl/.style={line width=0.9pt},
    pf/.style={line width=0.9pt},
    tick/.style={line width=0.9pt},
    darr/.style={dashed, line width=0.8pt, ->},
    dot/.style={circle, fill=black, inner sep=1.3pt},
  }

  \def\xO{0}
  \def\xOne{1}
  \def\xTwo{2}
  \def\xTmid{6}  
  \def\xT{9}     

  \draw[tl] (\xO,0) -- (\xT,0);

  \foreach \x/\lab in {\xO/0,\xOne/1,\xTwo/2,\xTmid/$t$,\xT/$T$}{
    \draw[tick] (\x,0.12) -- (\x,-0.12);
    \node[above] at (\x,0.18) {\lab};
  }
  \node[above] at (3.9,0.2) {$\cdots$};
  \node[above] at (7.5,0.2) {$\cdots$};

  \node[anchor=east] at (-0.25,-0.3) {portfolio};
  \node[anchor=east,color=gray] at (-0.25,-1.1) {\#1};
  \node[anchor=east,color=gray] at (-0.25,-2.00) {\#2};
  \node[anchor=east] at (-0.25,-3.30) {\#$t$};
  \node[anchor=east,color=gray] at (-0.25,-4.6) {\#$T$};

  \node[dot,color=gray] at (\xO,-1.10) {};
  \draw[pf,color=gray] (\xO,-1.10) -- (\xOne,-1.10);
  \node[above,color=gray] at (\xO+0.5,-1.10) {$u^{(1)}_{0}$};

  \draw[darr,color=gray] (\xOne,-1.10) -- (\xOne,-0.2);
  \node[right,color=gray] at (\xOne, -0.7) {$R_{1}$};

  \node[dot,color=gray] at (\xO,-2.00) {};
  \draw[pf,color=gray] (\xO,-2.00) -- (\xTwo,-2.00);
  \draw[tick,color=gray] (\xOne,-2.00+0.12) -- (\xOne,-2.00-0.12);
  \node[above,color=gray] at (\xO+0.5,-2.00) {$u^{(2)}_{0}$};
  \node[above,color=gray] at (\xOne+0.50,-2.00) {$u^{(2)}_{1}$};

  \draw[darr,color=gray] (\xTwo,-2.00) -- (\xTwo,-0.2);
  \node[right,color=gray] at (\xTwo, -1.10) {$R_{2}$};

  \node[color=gray] at (-0.55,-2.50) {$\vdots$};

  \node[dot] at (\xO,-3.30) {};
  \draw[pf] (\xO,-3.30) -- (\xTmid,-3.30);
  \draw[tick] (\xOne,-3.30+0.12) -- (\xOne,-3.30-0.12);
  \draw[tick] (\xOne+1.0,-3.30+0.12) -- (\xOne+1.0,-3.30-0.12);
  \draw[tick] (\xTmid-1.0,-3.30+0.12) -- (\xTmid-1.0,-3.30-0.12);
  \node[above] at (\xO+0.5,-3.30) {$u^{(t)}_{0}$};
  \node[above] at (\xOne+0.50,-3.30) {$u^{(t)}_{1}$};
  \node[above] at (\xTmid-0.5,-3.30) {$u^{(t)}_{t-1}$};
  \node[above] at (3.0,-3.30) {$\cdots$};

  \draw[darr] (\xTmid,-3.30) -- (\xTmid,-0.2);
  \node[right] at (\xTmid, -1.65) {$R_{t}$};

  \node[color=gray] at (-0.55,-3.80) {$\vdots$};

  \node[dot,color=gray] at (\xO,-4.6) {};
  \draw[pf,color=gray] (\xO,-4.6) -- (\xT,-4.6);
  \draw[tick,color=gray] (\xOne,-4.6+0.12) -- (\xOne,-4.6-0.12);
  \draw[tick,color=gray] (\xOne+1,-4.6+0.12) -- (\xOne+1,-4.6-0.12);
  \draw[tick,color=gray] (\xT-1,-4.6+0.12) -- (\xT-1,-4.6-0.12);
  \node[above,color=gray] at (\xO+0.5,-4.6) {$u^{(T)}_{0}$};
  \node[above,color=gray] at (\xOne+0.50,-4.6) {$u^{(T)}_{1}$};
  \node[above,color=gray] at (\xT-0.5,-4.6) {$u^{(T)}_{T-1}$};
  \node[above,color=gray] at (5.0,-4.6) {$\cdots$};

  \draw[darr,color=gray] (\xT,-4.6) -- (\xT,-0.2);
  \node[right,color=gray] at (\xT, -2.3) {$R_{T}$};

\end{tikzpicture}
        \caption{The RLOP method takes a forward, replication-based approach.}
        \label{fig:rlop-illustration}
    \end{minipage}

\end{figure}



The proposition below explains why the option price is increasing in the key parameters where higher risk aversion 
$\lambda$ and larger transaction friction both raise the price. We first prove monotonicity of the value function under any fixed policy, and then obtain monotonicity of the optimal value by maximization. The proof is in the supplementary materials.

\begin{prop}
For sufficiently large $\epsilon$ that appears in the linear transaction cost assumption $\text{TC}(\Delta u, S)=\epsilon|\Delta u| \,S$, the option price $C(S_0) := -\max_{\pi\in\mathbf{\Pi}} V_0^\pi$ is monotonically increasing in both $\lambda$ and $\epsilon$.
\label{prop:qlbs-monotone}
\end{prop}

\subsection{RLOP: A Forward Replication Learning Framework}

We propose Replication Learning of Option Pricing (RLOP) that takes a forward view.
The agent trades a self-financing portfolio and is rewarded based on how closely its terminal value matches the option payoff.
Compared with Deep Hedging (\cite{buehler2019deep}), which can embed a speculative component (\cite{franccois2025difference}), RLOP’s shortfall-probability objective encourages capital preservation and downside-aware hedging.
The reward shaping technique (\cite{sutton2018reinforcement,devlin2011theoretical}) suggests stacking an ensemble of maturities: along a path $\{S_t\}$ up to horizon $T$, the agent simultaneously manages portfolios $\Pi_t^{(i)}$ for expiries $i=1,\dots,T$, choosing hedge positions $u_t^{(i)}$ for all $t<i$.
This provides intermediate feedback and lets the policy learn from short horizons before scaling to the full maturity.
A formal definition of the RLOP problem is given as follows.

\begin{defn}
The transitional probability density from $X_{t}=\left(t,S_{t}\right)$ to $X_{t+1}$ is defined as
$
    p\left(X_t,R_{t+1}\big|X_{t+1},u_{t}^{\left(i\right)}\right)=  \rho\left(S_{t},S_{t+1}\right) \boldsymbol{1}_{t<i}
$
where $\rho$ is characterized by the discrete version geometric Brownian motion.
The associated reward function is $R_{i}=H\left(h\left(S_{i}\right),\Pi_{i}^{\left(i\right)}\right)$
where the penalty function $H$ measures how accurate
the portfolio value $\Pi_{i}^{\left(i\right)}$ replicates the option
payoff $h\left(S_{i}\right)$.
\end{defn}

In practice, we take $H(x,y)=-|x-y|$ or its squared variant, which directly penalizes terminal replication error.
The structure of the RLOP approach is illustrated in \cref{fig:rlop-illustration}.



\section{Learning Algorithms and Numerical Verification}

We parametrize the hedging policy in both QLBS and RLOP with neural networks and train it in a simulated environment.
The environment generates geometric-Brownian price paths with parameters $(r,\mu,\sigma,T)$.
At each time $t$, the agent observes the normalized state $(t,X_t)$, outputs a hedge position $a_t$, and receives rewards computed from \cref{def:qlbs}; performance is estimated by Monte Carlo rollouts.



We model the policy as a Gaussian $\pi=\mathcal N(\mu_\pi,\sigma_\pi)$, where $\mu_\pi$ and $\sigma_\pi$ are produced by a shared ResNet-style network (\cite{he2016deep}). A separate value network with the same architecture provides a learned baseline for variance reduction. Training uses REINFORCE with a baseline (\cite{williams1992simple,sutton2018reinforcement}), optimized by Adam at learning rate $10^{-4}$.



\begin{figure}[htbp]
  \centering
  \begin{minipage}[b]{0.49\textwidth}
    \centering
    \includegraphics[width=\textwidth]{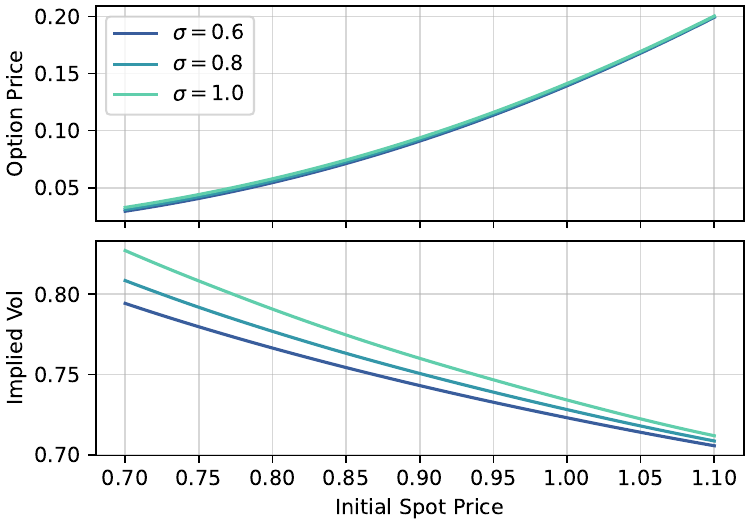}
    
  \end{minipage}
  \hfill
  \begin{minipage}[b]{0.49\textwidth}
    \centering
    \includegraphics[width=\textwidth]{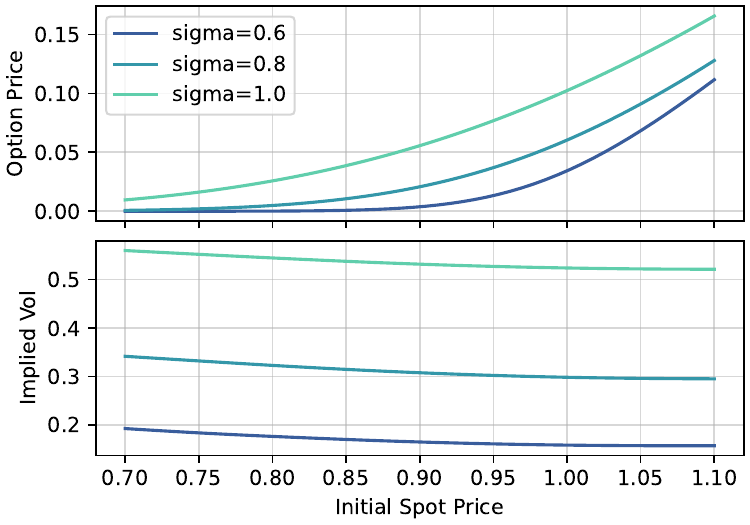}
  \end{minipage}
\caption{Price under Adaptive-QLBS model (left) and RLOP model (right) given different parameters of volatility. The common setup uses maturity $T=2$ months, strike $K=1$, interest rate $r=4\%$.}
\label{fig:sigma-qlbs-rlop}
\end{figure}

\begin{figure}[htbp]
  \centering
  \begin{minipage}[b]{0.3\textwidth}
    \centering
    \includegraphics[width=\textwidth]{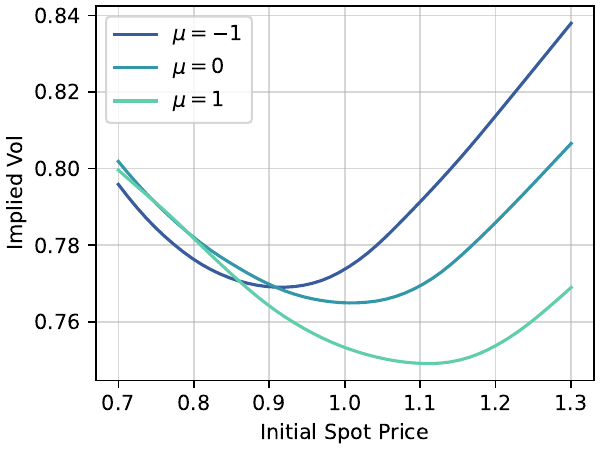}
  \end{minipage}
  \hfill
  \begin{minipage}[b]{0.3\textwidth}
    \centering
    \includegraphics[width=\textwidth]{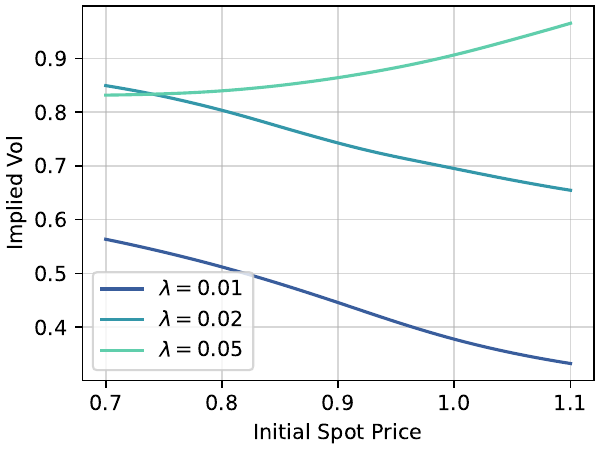}
  \end{minipage}
  \hfill
  \begin{minipage}[b]{0.3\textwidth}
    \centering
    \includegraphics[width=\textwidth]{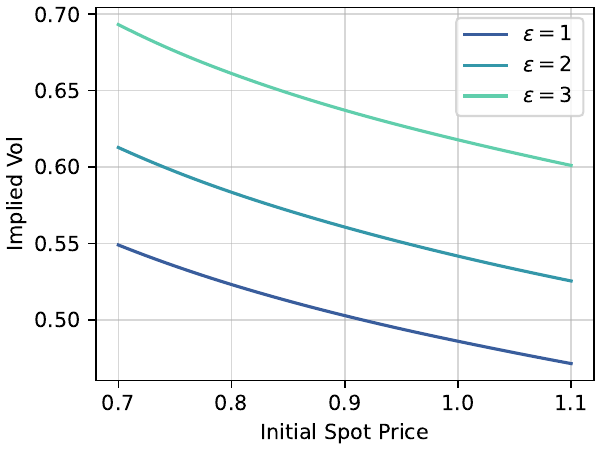}
  \end{minipage}
\caption{Price under Adaptive-QLBS model given different levels of hyperparameters: drift $\mu$ (left), risk aversion intensity $\lambda$ (middle), and friction $\epsilon$ (right).}
\label{fig:qq-qlbs}
\end{figure}

Finally, we verify the monotonicity in \cref{prop:qlbs-monotone} numerically. \Cref{fig:sigma-qlbs-rlop,fig:qq-qlbs} show that the learned prices vary consistently with $\sigma$, $\mu$, $\lambda$, and $\epsilon$, matching the theoretical trends and reproducing the implied-volatility skew patterns observed in practice.


\section{Empirical Evaluation on Market Data}

In this section we move from the simulated environments to market data. We study S\&P 500 and energy-sector ETF calls (SPY and XOP), and compare the two RL models (QLBS and RLOP) with three standard parametric baselines:
\cite{black1973pricing} (BS), \cite{merton1976discontinuous} jump-diffusion (JD), and the \cite{heston1993closed} stochastic-volatility model (SV).


\subsection{Data Description and Experimental Design}
\paragraph{Data}
We use daily snapshots of SPY and XOP call options from two non-overlapping quarters: 2020Q1 (COVID crash, high-volatility regime) and 2025Q2 (calmer regime). 
We construct synthetic European call prices $C^{\mathrm{mkt}}(K,\tau)$ and retain contracts with 3 to 70 calendar days to maturity.

\paragraph{Bucketing and moneyness}
Maturities are grouped into buckets centered at 14, 28, and 56 days. Within each bucket we define moneyness as $K/F$, where $F$ is the forward corresponding to the option's maturity. We focus on the 28-day bucket in the main analysis, and report the full maturity-by-moneyness breakdown in the appendix.

\paragraph{Daily calibration and fitting}
On each trading day and for each maturity bucket, we calibrate the parametric baselines to that day's cross-section by minimizing squared pricing errors. QLBS and RLOP are fit on the same slice using the procedure described in earlier sections. 

\paragraph{Evaluation}
We report two sets of outcomes.
(1) \textit{Static fit:} we compare model-implied volatilities with market implied volatilities using equal-day IVRMSE reported on a $10^3$ scale.
(2) \textit{Dynamic performance:} we evaluate discrete-time $\Delta$-hedging of a short call over the realized (``historical'') underlying path. Let
$
\Pi_T := V_T - (S_T-K)^+,
$
where $V_T$ is the terminal value of the self-financing hedging portfolio initialized with the option premium and including transaction costs. We report three metrics computed from $\Pi_T$:
\begin{enumerate}
\item Hedging RMSE: $\mathrm{RMSE}(\Pi_T):=\sqrt{\mathbb{E}[\Pi_T^2]}$, estimated by the root mean square of terminal $\Pi_T$ across hedges.
\item Average trading cost: cumulative proportional transaction costs incurred by hedge rebalancing over the horizon, reported per option.
\item Shortfall probability: $\mathbb{P}(\Pi_T<0)$, estimated in the historical backtest as the empirical frequency of negative terminal $\Pi_T$ across hedges.
\end{enumerate}

All reported numbers are equal-day averages: we compute each metric within a trading day and then average across days so that each day contributes one observation regardless of how many strikes are listed.

\subsection{Static Pricing Accuracy: Implied Volatility Fit}
Table~\ref{tab:ivrmse_28d_whole} reports equal-day IVRMSE for the 28-day maturity bucket (whole sample) for SPY and XOP in 2020Q1 and 2025Q2.
Three takeaways stand out.

\begin{table}[H]
\centering
\begin{tabular}{cc c c c c c c}
\toprule
Moneyness, $\tau$ & Period & Asset & BS & JD & SV & QLBS & RLOP \\
\midrule
\multirow{4}{*}{Whole sample, 28d}
& \multirow{2}{*}{2020Q1} & SPY & 76.50  & \textbf{17.62} & 42.08 & 102.65 & 82.47 \\
&                        & XOP & 120.18 & 91.63          & \textbf{88.34} & 109.92 & 188.07 \\
\cmidrule(lr){2-8}
& \multirow{2}{*}{2025Q2} & SPY & 127.95 & 94.90          & 74.02 & 92.54 & \textbf{73.43} \\
&                        & XOP & 106.10 & \textbf{64.83} & 73.30 & 111.23 & 163.62 \\
\bottomrule
\end{tabular}
\caption{Equal-day IVRMSE for the whole sample at $\tau=28$d. Lower is better; bold marks the best value within each row.}
\label{tab:ivrmse_28d_whole}
\end{table}

\begin{enumerate}
\item \textit{Parametric benchmarks fit the surface best, especially in stress.}
In SPY 2020Q1, JD substantially improves on BS and also beats SV, consistent with discontinuous moves mattering more in crash periods.

\item \textit{In calmer markets, the ranking compresses and can be asset-specific.}
In SPY 2025Q2, SV and RLOP deliver very similar IVRMSE, while JD remains clearly ahead of BS. For XOP 2025Q2, JD is best with SV close behind, suggesting that energy-sector options can exhibit different cross-sectional features than broad-index options even within the same quarter. 

\item \textit{RL models are not designed as implied-volatility interpolators, yet QLBS can be competitive in IV fit.}
Across the four asset–period pairs, QLBS usually sits between BS and the best parametric model, and is near JD in SPY 2025Q2. RLOP varies more, consistent with a frictional hedging objective rather than same-day IV fit.
\end{enumerate}

Full maturity and moneyness breakdowns are reported in
Appendix Tables~A.1--~A.3. 
The supplementary tables convey a broadly consistent qualitative picture: JD/SV are typically strongest in IVRMSE, while the hedging-oriented RL models are more variable across assets and regimes. 
We therefore turn next to realized-path $\Delta$-hedging results.

\subsection{Dynamic Hedging Performance}
We backtest each model’s discrete-time $\Delta$-hedging on the subsequent realized underlying path with proportional transaction costs, and summarize three desk-relevant diagnostics: RMSE of terminal hedging P\&L, average cumulative trading cost per option, and shortfall probability.

\paragraph{ATM benchmark (Table~\ref{tab:hedging_28d_atm})}

\begin{table}[t]
\centering
\small
\setlength{\tabcolsep}{5pt}
\renewcommand{\arraystretch}{1.15}

\begin{tabular}{c c c p{3.1cm} c c c c c}
\toprule
Moneyness, $\tau$ & Period & Asset & Metric & BS & JD & SV & QLBS & RLOP \\
\midrule
\multirow{12}{*}{ATM, 28d}
& \multirow{6}{*}{2020Q1}
& \multirow{3}{*}{SPY}
& Hedging RMSE & 5.88 & 7.03 & 8.15 & \textbf{5.62} & 6.39 \\
& & & Average trading cost & 2.21 & 2.62 & 3.52 & 2.09 & \textbf{1.95} \\
& & & Shortfall probability & 1.00 & 0.96 & 0.96 & 1.00 & \textbf{0.91} \\
\cmidrule(lr){3-9}
& & \multirow{3}{*}{XOP}
& Hedging RMSE & 0.40 & 0.42 & 0.45 & 0.54 & \textbf{0.36} \\
& & & Average trading cost & 0.10 & 0.10 & 0.11 & \textbf{0.09} & 0.11 \\
& & & Shortfall probability & 1.00 & 1.00 & 1.00 & 1.00 & \textbf{0.96} \\
\cmidrule(lr){2-9}
& \multirow{6}{*}{2025Q2}
& \multirow{3}{*}{SPY}
& Hedging RMSE & \textbf{6.07} & 6.56 & 7.57 & 7.01 & 6.70 \\
& & & Average trading cost & 3.58 & 3.77 & 4.69 & 3.91 & \textbf{3.09} \\
& & & Shortfall probability & 0.97 & \textbf{0.53} & 0.55 & 0.92 & 0.76 \\
\cmidrule(lr){3-9}
& & \multirow{3}{*}{XOP}
& Hedging RMSE & \textbf{1.38} & 1.60 & 1.69 & 1.47 & 1.52 \\
& & & Average trading cost & 0.85 & 0.87 & 1.05 & \textbf{0.80} & 0.83 \\
& & & Shortfall probability & 0.58 & 0.40 & 0.42 & 0.60 & \textbf{0.39} \\
\bottomrule
\end{tabular}

\caption{ATM ($\tau=28$d) delta-hedging performance under transaction costs.
Equal-day averages of hedging RMSE, average trading cost, and shortfall probability for a short call hedged over 28 days on SPY and XOP in 2020Q1 and 2025Q2. Shortfall probability is reported as a fraction in $[0,1]$; values of $1.00$ indicate that, within that cell, essentially all hedges end with $\Pi_T<0$. Lower is better; bold marks the best value within each row.}
\label{tab:hedging_28d_atm}
\end{table}

We first focus on ATM options as they are the standard benchmark for $\Delta$-hedging comparisons. Three patterns emerge.

\begin{enumerate}
\item \textit{QLBS prioritizes replication while keeping turnover near classical deltas.}
Empirically, in stress, it achieves the lowest hedging RMSE with trading costs that remain near the BS benchmark. In calmer markets, it can trade a small amount of RMSE optimality for execution efficiency.

\item \textit{RLOP behaves like a friction- and downside-aware policy.}
On SPY, it is the lowest-cost hedge in both regimes (an $\sim$14\% reduction vs BS in 2025Q2), while also reducing shortfall probability relative to BS in both quarters.
On XOP, it has the best RMSE in stress and the lowest shortfall in 2025Q2, with competitive costs.

\item \textit{RL policies remain well-behaved in the 2020Q1 stress test.}
Rather than collapsing under the COVID crash dynamics, both RL hedges remain economically sensible.
QLBS preserves replication performance without inflating turnover, while RLOP delivers a clear capital-efficiency and downside signature.
\end{enumerate}

\paragraph{Near-OTM stress test (Appendix Table~A.4)}
We repeat the same exercise for a mildly OTM target ($K/F=1.03$) at $\tau=28$d. The results reinforce RLOP’s downside-sensitive behavior under costs: in SPY 2025Q2 it lowers shortfall probability and trading cost with only a modest RMSE increase. For XOP, RLOP also improves RMSE and shortfall in 2020Q1, and both RL methods remain competitive in 2025Q2.

\medskip
\noindent To summarize, RL policies improve economically relevant quantities even when they do not dominate every metric. In particular, the parametric models can be strong on one dimension (e.g., RMSE in calmer regimes), yet the RL policies often achieve meaningful reductions in trading cost and/or shortfall probability.

\subsection{Static versus Dynamic Metrics: Robustness and Interpretation}

Our result shows that static surface fit and dynamic hedging performance are weakly aligned under discrete rebalancing and trading costs. The parametric benchmarks (especially JD and SV) often deliver the lowest IVRMSE, yet they do not consistently dominate realized hedging outcomes. This disconnect arises because IVRMSE is a one-day cross-sectional diagnostic, while hedging P\&L is shaped by realized price paths, re-hedging frequency, and turnover management under market frictions.

Therefore, the RL-based methods add a distinct and complementary value proposition. QLBS behaves like a cost-aware alternative to classical deltas: its hedging RMSE is typically in the same range as BS/JD/SV, while it often improves implementation cost and sometimes downside frequency, particularly outside the most extreme regions. RLOP prioritizes adverse outcome control. It frequently achieves lower shortfall probability and trading cost, even with higher dispersion of terminal hedging errors. By actively managing positions to strike an optimal balance between replication error (RMSE) and tail risk (shortfall probability), QLBS and RLOP computationally operationalizes the theoretical framework of \cite{follmer2000efficient}.

Finally, the qualitative interpretation is stable across regimes and horizons. The crisis window (2020Q1) and the calmer period (2025Q2) differ sharply in volatility conditions, yet the roles above remain intact. In the supplementary materials, we report the same ATM and near-OTM diagnostics for $\tau=14$d (Appendix Table~A.5 \&~A.6) and $\tau=56$d (Appendix Table~A.7 \&~A.8).
The conclusions remain broadly consistent: QLBS stays close to classical deltas in replication error while often improving cost efficiency, and RLOP continues to deliver meaningful reductions in loss frequency in many slices, especially in the near-OTM stress tests, where downside-sensitive behavior is most visible. 

\section{Conclusion}

This paper develops modified QLBS and novel RLOP environments for RL-based option pricing with transaction costs and risk aversion. Since pricing-model calibration and hedging performance are decoupled, this paper shifts the optimization objective toward shortfall probability to ensure tail-risk resilience to address the problem. We extend QLBS into a fully implemented algorithm, which acts as a cost-conscious hedge, and reduces shortfall probability and trading costs. RLOP prioritizes tail-risk resilience, accepting poorer IV fits for significant reductions in the probability and cost of extreme hedging losses and making it ideal for capital-constrained desks during market stress. Future work should incorporate computational complexity, path-dependent instruments, funding-spread jumps, and model risk to further improve pricing accuracy and execution efficiency.

\section*{Data Availability}
The option data used in this study are publicly available from the DoltHub repository \href{https://www.dolthub.com/repositories/post-no-preference/options}{post-no-preference/options}, which provides historical option records for SPY and XOP. The analyses in this paper use only these publicly accessible records, and no proprietary datasets were used. The data were accessed on Oct 15, 2025.

\section*{Funding Statement}
This research did not receive any specific grant from funding agencies in the public, commercial, or not-for-profit sectors.

\section*{Author contributions (CRediT)}
Ziheng Chen: Conceptualization; Formal analysis; Software (reinforcement learning training); Writing -- original draft (mathematical derivations and proofs).

Minxuan Hu: Software (empirical data fitting and hedging-cost calibration); Data curation; Validation; Writing -- original draft (tables, formatting, and interpretation).

Jiayu Yi: Writing -- original draft (introduction); Writing -- review \& editing.

Wenxi Sun: Writing – original draft (literature review); Writing -- review \& editing.

\setcounter{table}{0}
\appendix
\section{Tables on Market Data Fit Results}

\begin{table}[H]
\centering
\small
\begin{tabular}{cc c c c c c c}
\toprule
Moneyness, $\tau$ & Period & Asset & BS & JD & SV & QLBS & RLOP \\
\midrule
\multirow{4}{*}{Whole sample, 28d}
& \multirow{2}{*}{2020Q1} & SPY & 76.50 & \textbf{17.62} & 42.08 & 102.65 & 82.47 \\
&                        & XOP & 120.18 & 91.63 & \textbf{88.34} & 109.92 & 188.07 \\
\cmidrule(lr){2-8}
& \multirow{2}{*}{2025Q2} & SPY & 127.95 & 94.90 & 74.02 & 92.54 & \textbf{73.43} \\
&                        & XOP & 106.10 & \textbf{64.83} & 73.30 & 111.23 & 163.62 \\
\midrule

\multirow{4}{*}{Moneyness $<1$, 28d}
& \multirow{2}{*}{2020Q1} & SPY & 88.35 & \textbf{16.13} & 41.32 & 133.86 & 95.27 \\
&                        & XOP & 140.94 & \textbf{98.03} & 102.23 & 124.84 & 195.22 \\
\cmidrule(lr){2-8}
& \multirow{2}{*}{2025Q2} & SPY & 169.37 & 114.23 & 75.53 & 107.61 & \textbf{75.46} \\
&                        & XOP & 123.68 & \textbf{75.07} & 86.06 & 134.59 & 137.33 \\
\midrule

\multirow{4}{*}{Moneyness $>1$, 28d}
& \multirow{2}{*}{2020Q1} & SPY & 54.65 & \textbf{17.52} & 32.06 & 54.55 & 63.70 \\
&                        & XOP & 76.38 & 53.47 & \textbf{45.76} & 66.05 & 164.72 \\
\cmidrule(lr){2-8}
& \multirow{2}{*}{2025Q2} & SPY & \textbf{39.65} & 53.59 & 59.29 & 65.22 & 63.61 \\
&                        & XOP & 79.44 & \textbf{43.28} & 49.46 & 65.95 & 182.29 \\
\midrule

\multirow{4}{*}{Moneyness $>1.03$, 28d}
& \multirow{2}{*}{2020Q1} & SPY & 61.90 & \textbf{20.16} & 37.02 & 41.66 & 55.93 \\
&                        & XOP & 81.38 & 56.35 & \textbf{49.15} & 69.93 & 166.67 \\
\cmidrule(lr){2-8}
& \multirow{2}{*}{2025Q2} & SPY & \textbf{40.44} & 59.41 & 65.79 & 67.61 & 61.19 \\
&                        & XOP & 85.40 & \textbf{46.07} & 52.92 & 66.93 & 191.37 \\
\bottomrule
\end{tabular}
\caption{Equal-day IVRMSE for the $\tau=28$d maturity bucket across moneyness groups. Lower is better; bold marks the best value within each asset, period, and moneyness row.}
\label{tab:ivrmse_28d_by_moneyness}
\end{table}

\begin{table}[H]
\centering
\begin{tabular}{cc c c c c c c}
\toprule
Moneyness, $\tau$ & Period & Asset & BS & JD & SV & QLBS & RLOP \\
\midrule
\multirow{4}{*}{Whole sample, 14d}
& \multirow{2}{*}{2020Q1} & SPY & 91.49  & \textbf{55.27} & 74.74  & 126.16 & 107.38 \\
&                        & XOP & 187.68 & \textbf{163.92} & 188.65 & 203.85 & 249.38 \\
\cmidrule(lr){2-8}
& \multirow{2}{*}{2025Q2} & SPY & 164.35 & 135.95 & 108.14 & 118.28 & \textbf{94.86} \\
&                        & XOP & 151.73 & \textbf{95.57} & 123.06 & 151.55 & 201.01 \\
\bottomrule
\end{tabular}
\caption{Equal-day IVRMSE for the whole sample at $\tau=14$d. Lower is better; bold marks the best value within each asset and period row.}
\label{tab:ivrmse_14d_whole}
\end{table}

\begin{table}[H]
\centering
\begin{tabular}{cc c c c c c c}
\toprule
Moneyness, $\tau$ & Period & Asset & BS & JD & SV & QLBS & RLOP \\
\midrule
\multirow{4}{*}{Whole sample, 56d}
& \multirow{2}{*}{2020Q1} & SPY & 65.52  & \textbf{12.26} & 22.68  & 87.32  & 84.13 \\
&                        & XOP & 59.44  & 33.43 & \textbf{31.80} & 60.07  & 199.07 \\
\cmidrule(lr){2-8}
& \multirow{2}{*}{2025Q2} & SPY & 92.69  & 60.61 & \textbf{43.59} & 74.41  & 70.51 \\
&                        & XOP & 68.77  & 37.90 & \textbf{37.56} & 72.08  & 147.40 \\
\bottomrule
\end{tabular}
\caption{Equal-day IVRMSE for the whole sample at $\tau=56$d. Lower is better; bold marks the best value within each asset and period row.}
\label{tab:ivrmse_56d_whole}
\end{table}

\begin{table}[H]
\centering
\small
\setlength{\tabcolsep}{5pt}
\renewcommand{\arraystretch}{1.15}

\begin{tabular}{c c c p{3.1cm} c c c c c}
\toprule
Moneyness, $\tau$ & Period & Asset & Metric & BS & JD & SV & QLBS & RLOP \\
\midrule
\multirow{12}{*}{$K/F=1.03$, 28d}
& \multirow{6}{*}{2020Q1}
& \multirow{3}{*}{SPY}
& Hedging RMSE & \textbf{3.57} & 4.16 & 6.21 & 3.61 & 4.53 \\
& & & Average trading cost & 0.88 & 0.96 & 1.84 & \textbf{0.86} & 1.19 \\
& & & Shortfall probability & 1.00 & 1.00 & \textbf{0.96} & 1.00 & 1.00 \\
\cmidrule(lr){3-9}
& & \multirow{3}{*}{XOP}
& Hedging RMSE & 0.31 & 0.31 & 0.34 & 0.44 & \textbf{0.23} \\
& & & Average trading cost & \textbf{0.07} & 0.08 & 0.08 & \textbf{0.07} & \textbf{0.07} \\
& & & Shortfall probability & 1.00 & 1.00 & 1.00 & 1.00 & \textbf{0.78} \\
\cmidrule(lr){2-9}
& \multirow{6}{*}{2025Q2}
& \multirow{3}{*}{SPY}
& Hedging RMSE & \textbf{5.58} & 5.83 & 7.36 & 6.97 & 6.05 \\
& & & Average trading cost & 3.57 & 4.08 & 4.86 & 3.48 & \textbf{2.96} \\
& & & Shortfall probability & 0.84 & 0.76 & 0.84 & 0.87 & \textbf{0.55} \\
\cmidrule(lr){3-9}
& & \multirow{3}{*}{XOP}
& Hedging RMSE & 1.28 & 1.50 & 1.86 & \textbf{1.25} & 1.32 \\
& & & Average trading cost & 0.81 & 0.85 & 0.98 & \textbf{0.71} & 0.79 \\
& & & Shortfall probability & 0.53 & 0.51 & 0.49 & 0.39 & \textbf{0.33} \\
\bottomrule
\end{tabular}

\caption{OTM ($K/F=1.03$, $\tau=28$d) delta-hedging performance under transaction costs.
Equal-day averages of hedging RMSE, average trading cost, and shortfall probability for a short call hedged over 28 days on SPY and XOP in 2020Q1 and 2025Q2. Lower is better; bold marks the best value within each asset, period, and metric row.}
\label{tab:hedging_28d_otm}
\end{table}

\begin{table}[H]
\centering
\small
\setlength{\tabcolsep}{5pt}
\renewcommand{\arraystretch}{1.15}

\begin{tabular}{c c c p{3.1cm} c c c c c}
\toprule
Moneyness, $\tau$ & Period & Asset & Metric & BS & JD & SV & QLBS & RLOP \\
\midrule
\multirow{12}{*}{ATM, 14d}
& \multirow{6}{*}{2020Q1}
& \multirow{3}{*}{SPY}
& Hedging RMSE & 2.83 & 3.24 & 4.04 & 2.77 & \textbf{2.31} \\
& & & Average trading cost & 1.42 & 1.67 & 2.24 & 1.39 & \textbf{1.16} \\
& & & Shortfall probability & 1.00 & 0.97 & 0.97 & 1.00 & \textbf{0.94} \\
\cmidrule(lr){3-9}
& & \multirow{3}{*}{XOP}
& Hedging RMSE & 0.29 & 0.30 & 0.32 & 0.33 & \textbf{0.27} \\
& & & Average trading cost & \textbf{0.07} & \textbf{0.07} & 0.08 & \textbf{0.07} & \textbf{0.07} \\
& & & Shortfall probability & 1.00 & 1.00 & 1.00 & 1.00 & \textbf{0.96} \\
\cmidrule(lr){2-9}
& \multirow{6}{*}{2025Q2}
& \multirow{3}{*}{SPY}
& Hedging RMSE & 4.09 & 4.33 & 4.83 & 4.35 & \textbf{4.07} \\
& & & Average trading cost & 2.04 & 2.12 & 2.59 & 2.26 & \textbf{1.92} \\
& & & Shortfall probability & 0.97 & \textbf{0.67} & 0.72 & 0.91 & 0.72 \\
\cmidrule(lr){3-9}
& & \multirow{3}{*}{XOP}
& Hedging RMSE & \textbf{0.89} & 0.98 & 1.05 & 0.96 & 0.93 \\
& & & Average trading cost & 0.50 & 0.51 & 0.60 & 0.49 & \textbf{0.48} \\
& & & Shortfall probability & 0.62 & 0.51 & 0.52 & 0.57 & \textbf{0.43} \\
\bottomrule
\end{tabular}

\caption{ATM ($\tau=14$d) delta-hedging performance under transaction costs.
Equal-day averages of hedging RMSE, average trading cost, and shortfall probability for a short call hedged over 14 days on SPY and XOP in 2020Q1 and 2025Q2. Lower is better; bold marks the best value within each asset, period, and metric row.}
\label{tab:hedging_14d_atm}
\end{table}

\begin{table}[H]
\centering
\small
\setlength{\tabcolsep}{5pt}
\renewcommand{\arraystretch}{1.15}

\begin{tabular}{c c c p{3.1cm} c c c c c}
\toprule
Moneyness, $\tau$ & Period & Asset & Metric & BS & JD & SV & QLBS & RLOP \\
\midrule
\multirow{12}{*}{$K/F=1.03$, 14d}
& \multirow{6}{*}{2020Q1}
& \multirow{3}{*}{SPY}
& Hedging RMSE & \textbf{2.06} & 2.23 & 2.89 & 2.07 & 2.24 \\
& & & Average trading cost & 1.23 & 1.37 & 1.86 & \textbf{1.18} & 1.35 \\
& & & Shortfall probability & 1.00 & 1.00 & \textbf{0.98} & 1.00 & 1.00 \\
\cmidrule(lr){3-9}
& & \multirow{3}{*}{XOP}
& Hedging RMSE & 0.20 & 0.21 & 0.22 & 0.23 & \textbf{0.18} \\
& & & Average trading cost & \textbf{0.05} & 0.06 & 0.06 & \textbf{0.05} & \textbf{0.05} \\
& & & Shortfall probability & 1.00 & 1.00 & 1.00 & 1.00 & \textbf{0.88} \\
\cmidrule(lr){2-9}
& \multirow{6}{*}{2025Q2}
& \multirow{3}{*}{SPY}
& Hedging RMSE & \textbf{3.86} & 4.03 & 4.56 & 4.12 & 3.98 \\
& & & Average trading cost & 1.92 & 2.00 & 2.49 & 2.07 & \textbf{1.84} \\
& & & Shortfall probability & 0.88 & 0.74 & 0.76 & 0.85 & \textbf{0.72} \\
\cmidrule(lr){3-9}
& & \multirow{3}{*}{XOP}
& Hedging RMSE & \textbf{0.81} & 0.88 & 0.92 & 0.85 & 0.84 \\
& & & Average trading cost & 0.46 & 0.48 & 0.56 & \textbf{0.43} & 0.45 \\
& & & Shortfall probability & 0.56 & 0.55 & 0.55 & \textbf{0.44} & 0.48 \\
\bottomrule
\end{tabular}

\caption{OTM ($K/F=1.03$, $\tau=14$d) delta-hedging performance under transaction costs.
Equal-day averages of hedging RMSE, average trading cost, and shortfall probability for a short call hedged over 14 days on SPY and XOP in 2020Q1 and 2025Q2. Lower is better; bold marks the best value within each asset, period, and metric row.}
\label{tab:hedging_14d_otm}
\end{table}

\begin{table}[H]
\centering
\small
\setlength{\tabcolsep}{5pt}
\renewcommand{\arraystretch}{1.15}

\begin{tabular}{c c c p{3.1cm} c c c c c}
\toprule
Moneyness, $\tau$ & Period & Asset & Metric & BS & JD & SV & QLBS & RLOP \\
\midrule
\multirow{12}{*}{ATM, 56d}
& \multirow{6}{*}{2020Q1}
& \multirow{3}{*}{SPY}
& Hedging RMSE & \textbf{6.92} & 8.73 & 9.75 & 8.53 & 9.15 \\
& & & Average trading cost & \textbf{2.67} & 3.05 & 3.95 & 2.88 & 2.84 \\
& & & Shortfall probability & 0.85 & 0.79 & 0.76 & 0.82 & \textbf{0.62} \\
\cmidrule(lr){3-9}
& & \multirow{3}{*}{XOP}
& Hedging RMSE & 0.52 & 0.54 & 0.55 & 0.62 & \textbf{0.46} \\
& & & Average trading cost & 0.11 & 0.11 & 0.12 & \textbf{0.10} & 0.11 \\
& & & Shortfall probability & 0.95 & 0.95 & 0.95 & \textbf{0.86} & 0.95 \\
\cmidrule(lr){2-9}
& \multirow{6}{*}{2025Q2}
& \multirow{3}{*}{SPY}
& Hedging RMSE & 9.91 & 10.23 & 11.12 & 12.86 & \textbf{9.12} \\
& & & Average trading cost & 5.61 & 5.68 & 7.10 & 5.86 & \textbf{4.54} \\
& & & Shortfall probability & 0.46 & 0.45 & 0.44 & 0.60 & \textbf{0.28} \\
\cmidrule(lr){3-9}
& & \multirow{3}{*}{XOP}
& Hedging RMSE & \textbf{2.24} & 2.50 & 2.58 & 2.50 & 2.56 \\
& & & Average trading cost & 1.07 & 1.05 & 1.35 & 1.11 & \textbf{0.89} \\
& & & Shortfall probability & 0.51 & 0.44 & 0.46 & 0.52 & \textbf{0.38} \\
\bottomrule
\end{tabular}

\caption{ATM ($\tau=56$d) delta-hedging performance under transaction costs.
Equal-day averages of hedging RMSE, average trading cost, and shortfall probability for a short call hedged over 56 days on SPY and XOP in 2020Q1 and 2025Q2. Lower is better; bold marks the best value within each asset, period, and metric row.}
\label{tab:hedging_56d_atm}
\end{table}

\begin{table}[H]
\centering
\small
\setlength{\tabcolsep}{5pt}
\renewcommand{\arraystretch}{1.15}

\begin{tabular}{c c c p{3.1cm} c c c c c}
\toprule
Moneyness, $\tau$ & Period & Asset & Metric & BS & JD & SV & QLBS & RLOP \\
\midrule
\multirow{12}{*}{$K/F=1.03$, 56d}
& \multirow{6}{*}{2020Q1}
& \multirow{3}{*}{SPY}
& Hedging RMSE & \textbf{4.61} & 5.35 & 6.25 & 6.13 & 6.06 \\
& & & Average trading cost & 2.15 & 2.43 & 3.14 & \textbf{1.96} & 2.64 \\
& & & Shortfall probability & 0.80 & 0.80 & 0.80 & 0.80 & \textbf{0.20} \\
\cmidrule(lr){3-9}
& & \multirow{3}{*}{XOP}
& Hedging RMSE & 0.40 & 0.40 & 0.40 & 0.47 & \textbf{0.29} \\
& & & Average trading cost & 0.07 & 0.07 & 0.08 & \textbf{0.06} & \textbf{0.06} \\
& & & Shortfall probability & 1.00 & 1.00 & 1.00 & 1.00 & \textbf{0.70} \\
\cmidrule(lr){2-9}
& \multirow{6}{*}{2025Q2}
& \multirow{3}{*}{SPY}
& Hedging RMSE & \textbf{8.09} & \textbf{8.09} & 9.63 & 12.14 & 9.47 \\
& & & Average trading cost & 5.70 & 5.66 & 7.10 & 6.26 & \textbf{4.77} \\
& & & Shortfall probability & 0.46 & 0.46 & 0.46 & 0.60 & \textbf{0.24} \\
\cmidrule(lr){3-9}
& & \multirow{3}{*}{XOP}
& Hedging RMSE & \textbf{1.87} & 1.88 & 2.03 & 2.12 & 2.02 \\
& & & Average trading cost & \textbf{0.88} & 0.90 & 1.15 & 0.93 & 0.91 \\
& & & Shortfall probability & 0.51 & \textbf{0.49} & 0.51 & 0.56 & \textbf{0.49} \\
\bottomrule
\end{tabular}

\caption{OTM ($K/F=1.03$, $\tau=56$d) delta-hedging performance under transaction costs.
Equal-day averages of hedging RMSE, average trading cost, and shortfall probability for a short call hedged over 56 days on SPY and XOP in 2020Q1 and 2025Q2. Lower is better; bold marks the best value within each asset, period, and metric row.}
\label{tab:hedging_56d_otm}
\end{table}

\section{Proof for Proposition 1.}
\setcounter{prop}{0}
\begin{prop}
For sufficiently large $\epsilon$ that appears in the linear transaction cost assumption $\text{TC}(\Delta u, S)=\epsilon|\Delta u| \,S$, the option price $C(S_0) := -\max_{\pi\in\mathbf{\Pi}} V_0^\pi$ is monotonically increasing in both $\lambda$ and $\epsilon$.
\label{prop:qlbs-monotone}
\end{prop}
\begin{proof}
Given any policy $\pi$, the portfolio value at time $t$ may be written as
\begin{align}
\Pi_t = u_t  S_t + \epsilon \mathfrak{S}_t + \gamma^{T-t} \mathfrak{E}_t,
\quad \mathfrak{S}_t := \sum_{j=0}^{T-t-1} \gamma^{j+1} |\Delta u_{t+j}| S_{t+j+1}, 
\quad \mathfrak{E}_t := h(S_T) - u_{T-1}S_T
\label{eq:qlbs-proof-port-value}
\end{align}
under the self-financing condition
(eq. 1)
.
This expression is affine in $\epsilon$ and independent of $\lambda$.
Hence, the monotonicity of $V_t^\pi$ with respect to  $\lambda$ follows immediately from
(eq. 2)
, since the discounted risk term is always non-positive.
For $\epsilon\gg1$, the monotonicity in $\epsilon$ follows from the fact that $\text{Var}_t\Pi_{t}\left(X_{t}\right)$ is a quadratic function of $\epsilon$ with non-negative leading coefficient.

To establish that $C(S_0,\lambda)$ is increasing in $\lambda$, let $\pi(\lambda)$ denote the maximizer of $V_t^\pi$.
For $\lambda ' > \lambda$, we have
$V_t^{\pi(\lambda )}(S_t;\lambda) \ge V_t^{\pi(\lambda')}(S_t;\lambda)$
because $\pi(\lambda)$  maximizes the value function at level $\lambda$.
From the argument above, $V_t^{\pi(\lambda')}(S_t;\lambda') \le V_t^{\pi(\lambda')}(S_t;\lambda)$.
Combining these inequalities yields
\begin{align*}
C(S_0,\lambda') = -V_0^{\pi(\lambda')} (S_0;\lambda') \ge -V_0^{\pi(\lambda')} (S_0;\lambda) \ge -V_0^{\pi(\lambda)} (S_0;\lambda) = C(S_0,\lambda) .
\end{align*}
Thus, $C$ is monotone in $\lambda$.
The argument for its monotonicity in $\epsilon$ proceeds similarly.


\end{proof}

\bibliographystyle{elsarticle-harv}
\bibliography{ref}

\end{document}